\documentclass[11pt]{article}
\usepackage{authblk}
\usepackage{tikz}
\usepackage{array}
\usepackage{ytableau}
\usepackage{cite}
\usepackage{pifont}
\usepackage{times} 
\usepackage{amssymb,amsbsy}
\usepackage{amsmath,amscd}
\usepackage{graphics}
\usepackage{pstricks}
\def\C{{\mathbf{C}}}
\def\Z{{\mathbf{Z}}}
\def\l{{\lambda}}

\def\eq#1{(\ref{#1})}

\def\viz{{\slshape viz.}}
\newtheorem{exx}{Example}
\newenvironment{example}{\begin{exx}\par\normalfont}{\null\end{exx}}
\newcommand\cp[1]{{\mathbf{P}^{#1}}}
\begin{document}
\title{Determination of many-electron basis functions for a Quantum Hall 
ground state using Schur polynomials}	
\author[1,3]{Sudhansu S. Mandal\thanks{sudhansu[at]phy.iitkgp.ernet.in}}
\author[2]{Sutirtha Mukherjee\thanks{sutirtha[at]kias.re.kr}}
\author[3]{Koushik Ray\thanks{koushik[at]iacs.res.in}}
\affil[1]{Department of Physics and Centre for
Theoretical Studies, Indian Institute\\ of Technology, Kharagpur 721302,
India}
\affil[2]{Korea Institute for Advanced Study, Quantum Universe Center, Seoul 02455, Korea}
\affil[3]{Department of Theoretical Physics, Indian Association for the
Cultivation\\ of Science,Jadavpur, Kolkata 700032, India}

\maketitle
\begin{abstract}
\noindent 
A method for determining the ground state of a planar 
interacting many-electron system in a magnetic field perpendicular 
to the plane is described. The ground state wave-function is expressed as
a linear combination of a set of basis functions. Given only the flux
and the number of electrons describing an incompressible state, we use the
combinatorics of partitioning the flux among the electrons to derive the basis 
wave-functions as linear combinations of Schur polynomials. 
The procedure ensures that the basis wave-functions form 
representations of the angular momentum algebra. 
We exemplify the method by deriving the basis functions for 
the $\tfrac{5}{2}$ quantum Hall state with a few particles. 
\end{abstract}
\thispagestyle{empty}
Fractional quantum Hall effect at the filling factor $\tfrac{5}{2}$
is the first example in condensed matter systems wherein quasi-particles are 
predicted to obey a non-Abelian braiding statistics \cite{GWW,RG}. 
At this filling, 
the second Landau level of spin-up electrons is half-filled. Moore and Read
(MR) proposed a ground-state wave function \cite{MR} for this state with an
effective filling factor $\nu^* = 1/2$ which can be interpreted as a 
chiral $p$-wave superfluid state of composite fermions \cite{Jain} formed as a 
bound state of an electron and two flux quanta, each quantum being
$\phi_0=h c/e$. 
This fractional quantum Hall
state is also popularly known as the ``Pfaffian state" as the 
real-space representation of the chiral $p$-wave superfluid wave-function 
is a Pfaffian of a certain 
antisymmetric matrix depending on the relative positions 
of the particles. The edge of this state carries a neutral mode of 
Majorana fermions \cite{Wen}. 
It occurs for a flux $\Phi = 2N-3$ in units of $\phi_0$ in a 
system of $N$ electrons moving on the surface of a sphere\cite{hald1} of 
radius $\sqrt{\Phi/2}$.  On a disc, the orbitals occupied by $N$ electrons 
have an angular momentum less than or equal to $\Phi$. The quasi-holes of 
the MR state obey a non-Abelian braiding statistics \cite{GWW,RG} 
as the ground state 
of $2m$ quasi-holes is $2^{m-1}$-fold degenerate \cite{NW}. 
This wave-function is 
the exact ground state wave-function for a model three-body pseudo potential \cite{GWW}.
The MR wave-function 
has also been shown to be equivalent to the $\Z_2$ parafermion \cite{RR} 
wave-function. 
Jack polynomials characterized by occupation number configuration have been
introduced \cite{hald2} 
for the MR states. It naturally implements a squeezing rule 
in the configuration. However, no better alternative wave-function for the 
$\nu^* = 1/2$ state has been derived from this type of configuration so far.
A better wave-function is indeed necessary for understanding the ground 
state due for the Coulomb interaction.

In this article we propose a method to determine the basis wave-functions 
for a fractional quantum Hall state of $N$ electrons, 
with the maximum number of filled single-particle orbitals 
(same as the total number of flux quanta $\Phi$).
The ground state of the system is a suitable 
linear combination of these wave-functions. While the procedure is valid for 
arbitrary $\Phi$ and $N$, we consider, by way of example, the
``Pfaffian flux-shift" $\Phi = 2N-3$, for which the MR wave-function \cite{MR}
has been proposed as the ground state wave-function of the system. 
Numerical exact diagonalization studies \cite{Rezayi}, however, 
indicate that the MR 
wave-function is not the veritable ground state wave-function for the 
Coulomb interaction. The basis wave-functions obtained using our prescription 
should provide an accurate ground state.

Our method, being combinatorial in nature, has the advantage of being
conceptually simple, if computationally demanding. 
The assumptions made are
few and most natural, rendering the method very general. We obtain the basis
wave-functions solely from the knowledge of the integral flux and 
the number of electrons.

{\it Construction:} Let us consider a collection of $N$ electrons  
on the complex plane $\C$, at positions
$\{z_i|z_i\in\C,i=1,\cdots,N\}$ in a magnetic field with a given total 
integral flux $\Phi$ in units of the flux 
quantum $\phi_0$. In other words, $\Phi$ is a given, arbitrary, 
positive integer. 
The wave-function of this many-particle system is sought in the form 
$\Psi(z_1,z_2,\cdots, z_N)e^{-\tfrac{1}{4}\sum_k |z_k|^2}$, where $\Psi$ is 
a polynomial in the coordinate ring $\C[z_1,z_2,\cdots, z_N]$. 
We assume that the physical quantities derived using this wave funciton remain
unaltered as the electrons are shuffled. 
Assuming further that $\Phi$ is distributed as flux lines
between different pairs of electrons formed for a single electron, so that every unit of flux is divided
between two electrons of the pair in two moieties, and every pair of 
electron feels at least one unit of flux, the total angular momentum 
of the collection is \cite{hald1}
\begin{equation} 
\label{L:def}
L=N\Phi/2. 
\end{equation} 
Requiring the wave-function to furnish a representation of the angular momentum
algebra implies that the polynomial $\Psi$ is homogeneous of degree
$L$. 
Each electron feels this flux. The maximal index of any $z_i$ in $\Psi$ 
is thus $\Phi$. 
Let us define $z_{ij}=z_i-z_j$.  The number of such variables is
\begin{equation}
\label{E:def} 
E=\binom{N}{2},
\end{equation} 
the number of ways $N$ objects can be paired.
Since two electrons are not allowed to be at the same
position, the wave-function is supposed to vanish at $z_{ij}=0$, for every
pair of $i$ and $j$. This is ensured by assuming that the polynomial $\Psi$ is
antisymmetric under the exchange of $z_i$ and $z_j$. Without any loss of
generality we assume it to be of the form
\begin{equation}
\label{Psi:def}
\Psi(z_1,\cdots z_N) =  \Delta(z_1,\cdots,z_N) S^{(N)}(z_1,\cdots,z_N),
\end{equation} 
where $\Delta(z_1,\cdots,z_N)=\prod\limits_{1\leqslant i<j\leqslant N} z_{ij}$
is the Vandermonde polynomial. Since any physical quantity
is evaluated in terms of the modulus of the wave-function, assuming the
wave-function to be
antisymmetric under the exchange of electrons is consistent with the
invariance under shuffling.
Each $z_{ij}$ occurs in the Vandermonde polynomial as a factor only once.
Hence $S^{(N)}$ is a symmetric homogeneous 
polynomial in $\C[z_1,z_2,\cdots,z_N]$ 
containing each  $z_i$ with maximal index
\begin{equation}
\label{delta:def}
\delta=\Phi-(N-1), 
\end{equation} 
and degree
\begin{equation}
\begin{split}
\label{D:def}
D&=L-E\\&=\frac{N}{2}\delta.
\end{split}
\end{equation} 
From now on we shall deal mostly with $S^{(N)}$. 

We propose a systematic way of deriving the symmetric polynomial. 
It is useful to describe the system in terms of graphs consisting in
vertices and edges. We need to consider closed graphs, the ones in which
every edge connects a pair of vertices. The positions $z_i$ 
of the $N$ electrons are associated to the vertices of a graph $\mathcal{G}$.
The edges are associated with $z_{ij}$, $i\neq j$. 
The edges are assigned weights, such that $z_{ij}$ is given a weight
$a_{ij}$. The matrix $A=\{(a_{ij})| i,j=1,2,\cdots,N\}$ is referred to as
the  adjacency matrix of $\mathcal G$.
Each vertex is $\delta$-valent,  with the total weight of the edges attached to
it equal to $\delta$. We therefore consider a directed $\delta$-regular graph 
$\mathcal{G}=\big(\{z_i\},\{z_{ij}\}; (a_{ij})|i,j=1,2,\cdots,N\big)$ 
without loops.
Let us note that since the requirement of every pair of electrons being
related by at least one unit of flux has been taken care of by the
Vandermonde polynomial, the graph $\mathcal G$, which is but associated to the
symmetric polynomial $S^{(N)}$, need not be connected.
To the graph $\mathcal{G}$ we associate a \emph{graph monomial} 
\begin{equation}
\label{PG:def}
P_{\mathcal G}(z_1,z_2,\cdots,z_N) = 
\prod\limits_{1\leqslant i<j\leqslant N} \big(z_{ij}\big)^{a_{ij}}.
\end{equation} 
We identify the symmetric polynomials $S^{(N)}$ in \eq{Psi:def} 
with the symmetric graph polynomial obtained by 
symmetrizing the graph monomial,
\begin{equation}
\label{SN1}
S^{(N)}(z_1,z_2,\cdots,z_N) = \sum_{\varpi\in \mathfrak{S}_N} 
P_{\mathcal G}(z_{\varpi(1)},z_{\varpi(2)},\cdots,z_{\varpi(N)}),
\end{equation} 
where $\mathfrak{S}_N$ denotes the permutation group of order $N$.
We determine the graph monomial by  partitioning the total flux.
The flux $\Phi$ associated to each of the $N$ electrons is shared with every 
other electron, by assumption. We thus consider partitions of $\Phi$
into $N-1$ parts. The number of such partitions is obtained
as the coefficient of $y^{N-1}x^{\Phi}$ in the series expansion 
of the generating function
\begin{equation}
Z=\prod_{i=1}^{\infty}\frac{1}{1-yx^i}.
\end{equation} 
For a given partition $\lambda\vdash\Phi$, 
$\lambda=(\l_1,\l_2,\cdots,\l_{N-1})$ of length $|\lambda|=N-1$,
writing the wave-function \eq{Psi:def} as the antisymmetrized form of
\begin{equation}
\begin{split}
\label{Psi:partition}
z^{\l_1}_{12}z^{\l_2}_{13}\cdots 
z^{\l_{N-1}}_{1N}z^{\l_3}_{23}z^{\l_4}_{24}\cdots 
z^{\l_{N-1}}_{2(N-1)}z^{\l_{2}}_{2N}
z^{\l_5}_{34}z^{\l_6}_{35} &\cdots
\end{split}
\end{equation} 
ensures that the maximum index of each $z_i$ is 
$\sum_{i=1}^{|\lambda|}\lambda_i=\Phi$, while the degree 
equals the total angular momentum $L$ in the polynomial.
Taking out the Vandermonde polynomial, as in \eq{Psi:def}, is tantamount to
reducing the index  of each $z_{ij}$ by unity. Thus, given a partition
$\l=(\l_1,\l_2,\cdots,\l_{N-1})$, the reduced entries 
$\l_i-1$ determine the adjacency matrix of $\mathcal G$.
Taking this into account, the graph monomial is written as
\begin{equation}
\begin{split}
\label{PG1}
P_{\mathcal G}(z_1,\cdots,z_N) = z^{\l_1-1}_{12}z^{\l_2-1}_{13}\cdots 
z^{\l_{N-1}-1}_{1N}z^{\l_3-1}_{23}z^{\l_4-1}_{24}\cdots 
z^{\l_{N-1}-1}_{2(N-1)}z^{\l_2-1}_{2N}
z^{\l_5-1}_{34}z^{\l_6-1}_{35} \cdots.
\end{split}
\end{equation} 
The partitions can be obtained using \verb|Mathematica|, for example. We
present a Sudoku-like method to obtain the adjacency matrix $A$ from a
given partition. We demonstrate this for the case of $N=10$. 
Let us begin with a table of size $10\times 10$. We consider a partition
$(\l_1,\l_2,\cdots,\l_9)$. These numbers are to be arranged in the table
obeying the following rules. First, the diagonal boxes should be empty. 
Secondly, no $\lambda$ may be repeated along a row or a column. 
Thirdly, the sum of every row and every column must equal $\sum\lambda_i$. 
To start with we fill the first row and the first column of the table 
with $\l_1$--$\l_9$ as shown, leaving the $(1,1)$ entry empty. The
successive rows are then filled with $\l$'s keeping the first column intact,
which necessitates a protrusion of the table on the left. 
At this stage we label the diagonal entries till the last label, \viz
$\l_9$, stops before the diagonal. This process fills the table from the top
left up to the skew-diagonal. 
We then fix the entries in the white block, 
except for the last row. Then the protruded blocks as well as the
diagonal entries are shifted inside the table as indicated by the colored
blocks. The last row is then filled up uniquely. 
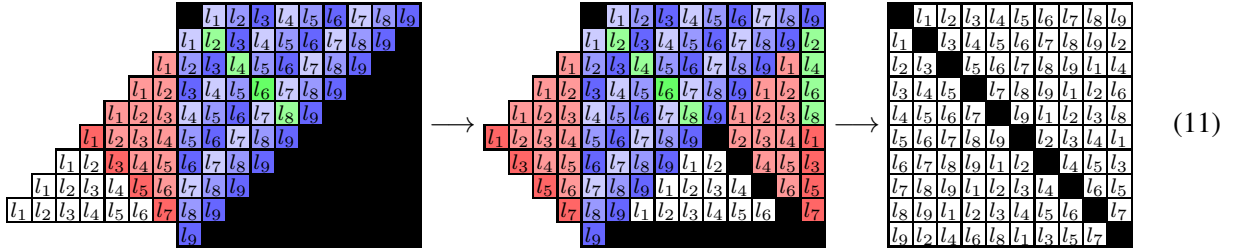
\begin{equation}
\label{yytab}
\ytableausetup{boxsize=.8em,centertableaux}
\ytableaushort[\scriptstyle\l_]{
\none\none\none\none\none\none\none
{*(black)0}{*(blue!20)1}{*(blue!40)2}{*(blue!60)3}{*(blue!20)4}{*(blue!40)5}
{*(blue!60)6}{*(blue!20)7}{*(blue!40)8}{*(blue!60)9},
\none\none\none\none\none\none\none{*(blue!20)1}{*(green!40)2}{*(blue!60)3}{*(blue!20)4}
{*(blue!40)5}{*(blue!60)6}{*(blue!20)7}{*(blue!40)8}{*(blue!60)9}{*(black)2},
\none\none\none\none\none\none{*(red!40)1}{*(blue!40)2}{*(blue!60)3}{*(green!40)4}
{*(blue!40)5}{*(blue!60)6}
{*(blue!20)7}{*(blue!40)8}{*(blue!60)9}{*(black)1}{*(black)4},
\none\none\none\none\none{*(red!40)1}{*(red!40)2}{*(blue!60)3}{*(blue!20)4}{*(blue!40)5}{*(green!60)6}{*(blue!20)7}
{*(blue!40)8}{*(blue!60)9}{*(black)1}{*(black)2}{*(black)6},
\none\none\none\none
{*(red!40)1}{*(red!40)2}{*(red!40)3}{*(blue!20)4}{*(blue!40)5}{*(blue!60)6}{*(blue!20)7}{*(green!40)8}{*(blue!60)9}
{*(black)1}{*(black)2}{*(black)3}{*(black)8},
\none\none\none
{*(red!60)1}{*(red!40)2}{*(red!40)3}{*(red!40)4}{*(blue!40)5}{*(blue!60)6}{*(blue!20)7}{*(blue!40)8}{*(blue!60)9}
{*(black)0}{*(black)2}{*(black)3}{*(black)4}{*(black)1},
\none\none
12{*(red!60)3}{*(red!40)4}{*(red!40)5}
{*(blue!60)6}{*(blue!20)7}{*(blue!40)8}{*(blue!60)9}{*(black)1}{*(black)2}{*(black)0}
{*(black)4}{*(black)5}{*(black)3},
\none1234{*(red!60)5}{*(red!40)6}
{*(blue!20)7}{*(blue!40)8}{*(blue!60)9}
{*(black)1}{*(black)2}{*(black)3}{*(black)4}
{*(black)0}{*(black)6}{*(black)5},
123456{*(red!60)7}
{*(blue!40)8}{*(blue!60)9}
{*(black)0}{*(black)7}
{*(black)0}{*(black)7}
{*(black)0}{*(black)7}
{*(black)0}{*(black)7},
\none\none\none\none\none\none\none
{*(blue!60)9}
{*(black)0}
{*(black)0}
{*(black)0}
{*(black)0}
{*(black)0}
{*(black)0}
{*(black)0}
{*(black)0}
{*(black)0}
}
\longrightarrow
\ytableausetup{boxsize=.8em,centertableaux}
\ytableaushort[\scriptstyle\l_]{
\none\none\none\none
{*(black)0}{*(blue!20)1}{*(blue!40)2}{*(blue!60)3}{*(blue!20)4}{*(blue!40)5}
{*(blue!60)6}{*(blue!20)7}{*(blue!40)8}{*(blue!60)9},
\none\none\none\none{*(blue!20)1}{*(green!40)2}{*(blue!60)3}{*(blue!20)4}
{*(blue!40)5}{*(blue!60)6}{*(blue!20)7}{*(blue!40)8}{*(blue!60)9}{*(green!40)2},
\none\none\none{*(red!40)1}{*(blue!40)2}{*(blue!60)3}{*(green!40)4}
{*(blue!40)5}{*(blue!60)6}
{*(blue!20)7}{*(blue!40)8}{*(blue!60)9}{*(red!40)1}{*(green!40)4},
\none\none{*(red!40)1}{*(red!40)2}{*(blue!60)3}{*(blue!20)4}{*(blue!40)5}{*(green!60)6}{*(blue!20)7}
{*(blue!40)8}{*(blue!60)9}{*(red!40)1}{*(red!40)2}{*(green!40)6},
\none{*(red!40)1}{*(red!40)2}{*(red!40)3}{*(blue!20)4}{*(blue!40)5}{*(blue!60)6}{*(blue!20)7}{*(green!40)8}{*(blue!60)9}
{*(red!40)1}{*(red!40)2}{*(red!40)3}{*(green!40)8},
{*(red!60)1}{*(red!40)2}{*(red!40)3}{*(red!40)4}{*(blue!40)5}{*(blue!60)6}{*(blue!20)7}{*(blue!40)8}{*(blue!60)9}
{*(black)0}{*(red!40)2}{*(red!40)3}{*(red!40)4}{*(red!60)1},
\none{*(red!60)3}{*(red!40)4}{*(red!40)5}
{*(blue!60)6}{*(blue!20)7}{*(blue!40)8}{*(blue!60)9}12{*(black)0}
{*(red!40)4}{*(red!40)5}{*(red!60)3},
\none\none{*(red!60)5}{*(red!40)6}
{*(blue!20)7}{*(blue!40)8}{*(blue!60)9}1234{*(black)0}{*(red!40)6}{*(red!60)5},
\none\none\none{*(red!60)7}
{*(blue!40)8}{*(blue!60)9}123456{*(black)0}{*(red!60)7},
\none\none\none\none
{*(blue!60)9}
{*(black)0}
{*(black)0}
{*(black)0}
{*(black)0}
{*(black)0}
{*(black)0}
{*(black)0}
{*(black)0}
{*(black)0}
}
\longrightarrow
\ytableaushort[\scriptstyle\l_]{
{*(black)0}123456789,
1{*(black)2}34567892,
23{*(black)4}5678914,
345{*(black)6}789126,
4567{*(black)8}91238,
56789{*(black)0}2341,
678912{*(black)0}453,
7891234{*(black)0}65,
89123456{*(black)0}7,
924681357{*(black)0}
}
\end{equation} 
Subtracting unity from each entry the upper triangular part of the table
specifies  the adjacency matrix
\begin{equation}
A = \begin{pmatrix}
0&\l_1' 
& \l_2'
& \l_3'
& \l_4'
& \l_5'
& \l_6'
& \l_7'
& \l_8'
& \l_9'\\
&0
& \l_3'
& \l_4'
& \l_5'
& \l_6'
& \l_7'
& \l_8'
& \l_9'
& \l_2'\\
&&0
& \l_5'
& \l_6'
& \l_7'
& \l_8'
& \l_9'
& \l_1'
& \l_4'\\
&&&0
& \l_7'
& \l_8'
& \l_9'
& \l_1'
& \l_2'
& \l_6'\\
&&&&0
& \l_9'
& \l_1'
& \l_2'
& \l_3'
& \l_8'\\
&&&&&0
& \l_2'
& \l_3'
& \l_4'
& \l_1'\\
&&&&&&0
& \l_4'
& \l_5'
& \l_3'\\
&&&&&&&0
& \l_6'
& \l_5'\\
&&&&&&&&0
& \l_7'\\
&&&&&&&&&0
\end{pmatrix},
\end{equation}
where we denoted $\l_i'=\l_i-1$. 
Using this in \eq{PG:def} yields \eq{PG1}.

Writing the wave-function in terms of the graph monomial serves purposes
beyond semantic and book-keeping ones. The symmetric graph polynomial
$S^{(N)}$ can be expressed as a polynomial in terms of the elementary 
symmetric polynomials 
$(e_1,e_2,\cdots,e_N)$ \cite{Macdonald}. For $\delta$-regular graphs which we
consider, the degree of the graph symmetric polynomial in terms of
the elementary symmetric polynomials is $\delta$. Homogenizing it 
(in terms of the symmetric polynomials) one obtains 
a homogeneous polynomial of degree $D$ which is $SL(2,\C)$-invariant if
non-vanishing \cite{GS,vakil}. At the level of algebra 
the $SL(2,\C)$ may be identified with the angular momentum algebra 
\cite{Li} with total angular momentum $D$, generated by 
\begin{equation} 
J_+ = -\sum_{i=1}^N z_i^2\frac{\partial}{\partial
z_i}+\delta z_i,\quad
J_0 = \sum_{i=1}^N z_i\frac{\partial}{\partial z_i}-D,\quad
J_- = \sum_{i=1}^N \frac{\partial}{\partial z_i},
\end{equation} 
satisfying 
\begin{equation}
[J_0,J_{\pm}] = J_{\pm},\quad [J_+,J_-] = 2J_0.
\end{equation} 
Thus the construction of the wave-function guarantees that it is a
representation of the angular momentum algebra. 

This method yields as many symmetric polynomials $S^{(N)}$, by \eq{SN1},
as there are partitions $\l\vdash\Phi$, starting with a graph polynomial 
written using \eq{yytab} for each partition. 
However, it turns out that these are not 
linearly independent. The expressions of the
symmetric polynomials in terms of the elementary symmetric functions is not
particularly adept for finding the independent ones. It is more useful to
express the symmetric polynomials
as linear combinations of either the monomial symmetric polynomials  as well
as Schur polynomials \cite{Macdonald}. 
These can be achieved using combinations of 
codes in \verb|Mathematica| and \verb|Macaulay2|\cite{M2}.
At this point
let us recall that to a partition $\lambda=(\lambda_1,\cdots,\lambda_n)$, with
$\lambda_1\geqslant\lambda_2\geqslant\cdots\geqslant\lambda_n$,
is associated a Schur polynomial given by the Jacobi-Trudi formula
\cite{stone}
\begin{equation}
s_{(\lambda)} = \frac{\det\left(z_i^{\lambda_j+n-j}\right)}{ 
\det\left(z_j^{n-i}\right)}.
\end{equation}  
However, not all the Schur polynomials associated to partitions
$\l\vdash\Phi$ may appear in $S^{(N)}$. Let us recall that while $\Psi$ is
associated to the partitions $\l\vdash\Phi$, the symmetric polynomials are
associated to $\l_i-1$, with different degrees. Indeed, as specified before,
they are related to the 
generating function of the number of partitions of an integer $D$ into
at most $N$ parts each of size at most $\delta$, namely,
\begin{equation}
\label{gen:formula}
\mathcal{P}_N(\leqslant \delta,x) 
= \prod_{i=1}^N \frac{1-x^{\delta+N-i+1}}{1-x^i}.
\end{equation}
The maximal number of Schur polynomials whose linear combination determines
$S^{(N)}$ is thus obtained as the coefficient of $x^D$ in the series
expansion of $\mathcal{P}_N(\leqslant \delta,x)$, using the expressions
\eq{delta:def} and \eq{D:def}.

We shall now exemplify the procedure
for the $\tfrac{5}{2}$ quantum hall state in a few examples. The computations
become rather demanding in terms of computer memory as the number of
electrons is increased.
 
{\it Examples:} Let us work out some examples using the procedure described 
above. We shall consider the cases $N=4,6,8$. 
We restrict attention to the $\tfrac{5}{2}$-state. 
Hence we have $\Phi = 2N-3$. Thus, only the knowledge of $N$ is necessary to 
derive the symmetric polynomials. Partitions $\l\vdash\Phi$
for these cases are given in Table~\ref{tab1}. The Young tableaux are all
of height $N-1$. 
\begin{table}[h]
\centering
\ytableausetup{smalltableaux,boxsize=5pt}
\begin{tabular}{cccl}
\hline
\hline
$N$ & $\Phi=2N-3$ & {\small Number} & {\small Partitions}\\
&&{\small of partitions}&\\
\hline
& & & \\
4 &5& 2 & \ydiagram{3,1,1}\,\,\,\ydiagram{2,2,1}\\
& & & \\
6 & 9& 5 &
\ydiagram{5,1,1,1,1,}\,\,\,\ydiagram{4,2,1,1,1}\,\,\,\ydiagram{3,3,1,1,1}\,\,\,
\ydiagram{3,2,2,1,1}\,\,\,\ydiagram{2,2,2,2,1}
\\
& & & \\
8& 13& 11 &\ydiagram{7,1,1,1,1,1,1}\,\,\,\ydiagram{6,2,1,1,1,1,1}\,\,\,
\ydiagram{5,3,1,1,1,1,1}\,\,\,
\ydiagram{5,2,2,1,1,1,1}\,\,\,
\ydiagram{4,4,1,1,1,1,1}\,\,\,
\ydiagram{4,3,2,1,1,1,1}\,\,\,
\\
& & & \\
&&&\ydiagram{4,2,2,2,1,1,1}\,\,\,
\ydiagram{3,3,3,1,1,1,1}\,\,\,\ydiagram{3,3,2,2,1,1,1}\,\,\,
\ydiagram{3,2,2,2,2,1,1}\,\,\,\ydiagram{2,2,2,2,2,2,1}
\end{tabular}
\caption{Number of electrons and partition of flux}
\label{tab1}
\end{table}
\begin{example}
Let us find the basis functions for $N=4$ electrons. The two partitions of
$\Phi=5$ are
$(3,1,1)$ and $(2,2,1)$, as shown in Table~\ref{tab1}. For the first case the
graph monomial and the graph are
\begin{alignat}{3}
P_{{\mathcal G}_1}=z_{12}^2z_{34}^2,\quad & {\mathcal G}_1 : &
\begin{tikzpicture}[baseline={([yshift=-.5ex]current bounding box.center)}]
\node (v1) {$1$};
\node [right of = v1] (v4) {$2$};
\node [below of = v1] (v3) {$3$};
\node [right of = v3] (v2) {$4$};
\draw[transform canvas={yshift=-1.5pt}] (v1) -- (v4);
\draw[transform canvas={yshift=1.5pt}] (v1) -- (v4);
\draw[transform canvas={yshift=-1.5pt}] (v3) -- (v2);
\draw[transform canvas={yshift=1.5pt}] (v3) -- (v2);
\end{tikzpicture}
\end{alignat}
The graphs are directed with edges $i\longrightarrow j$, for $i<j$. We shall
not indicate the directions explicitly in this article to avoid cluttering. 
For the second instance we have
\begin{alignat}{2}
P_{{\mathcal G}_2}=z_{12}z_{13}z_{24}z_{34} ,
\quad & {\mathcal G}_2: &
\begin{tikzpicture}[baseline={([yshift=-.5ex]current bounding box.center)}]
\node (v1) {$1$};
\node [right of = v1] (v4) {$2$};
\node [below of = v1] (v3) {$3$};
\node [right of = v3] (v2) {$4$};
\draw (v1) -- (v4) -- (v2) -- (v3) -- (v1);
\end{tikzpicture}
\end{alignat}
Let us emphasize that the graph monomial is directly obtained from the
partitions using \eq{PG1}, which in turn can be used to draw the graphs. 
Moreover, the entries of the adjacency matrix is depicted as the number of
lines between vertices.
The resulting symmetric polynomials turn out to be related,
\begin{equation}
S^{(4)}_1=2S^{(4)}_2=4e_4-e_1e_3+\frac{1}{3}e_2^2,
\end{equation} 
in terms of the elementary symmetric polynomials \cite{Macdonald}, 
\begin{equation} 
e_1 = z_1+z_2+z_3+z_4,\qquad e_2=z_1z_2+z_1z_3+...
\end{equation} 
Hence for four electrons there is a single symmetric polynomial which
determines $\Psi$ in \eq{Psi:def}. Let us point out that the degree of the
symmetric polynomial above in terms of the elementary symmetric polynomials
is $\delta=2$, hence the graphs are $2$-regular. 
The symmetric polynomial can also be expressed in terms of the monomial 
symmetric polynomials, $m$ as well as the three Schur polynomials $s$ as
\begin{equation}
\ytableausetup{boxsize=3pt}
\begin{split}
S^{(4)}_2 &= \frac{1}{3}
m_{\ydiagram{2,2}}-\frac{1}{3}m_{\ydiagram{2,1,1}}+2m_{\ydiagram{1,1,1,1}},\\
&=\frac{1}{3}\Big(
{s}_{(2,2)}-2 {s}_{(2,1,1)}+10 {s}_{(1,1,1,1)} 
\Big),
\end{split}
\end{equation}  
where we indicated the partitions of $4$ as subscripts in both cases. 
The maximal number of allowed 
Schur polynomials obtained from \eq{gen:formula} with
$N=4$, $\delta=2$ and $D=4$ is indeed $3$.
The symmetric polynomial is expressed in terms of
elementary and monomial symmetric polynomials using \verb|Mathematica|.
The former is then used to express
it in terms of Schur polynomials using \verb|Macaulay2|.
\end{example}
\begin{example}
Next, we consider $N=6$ particles. The partitions shown in Table~\ref{tab1}
yield the graph monomials
\begin{gather}
P_{{\mathcal G}_1}=
z^4_{12}z^4_{35}z^4_{46}\\
P_{{\mathcal G}_2}=
z_{12}^3z_{13}z_{26}z_{35}^3z_{45}z^3_{46}\\
P_{{\mathcal G}_3}=
z^2_{12}z^2_{13}z^2_{26}z^2_{35}z^2_{45}z^2_{46}\\
P_{{\mathcal G}_4}=
z_{12}^2z_{13}z_{14}z_{23}z_{26}z_{35}^2z_{45}z_{46}^2z_{56}\\
P_{{\mathcal G}_5}=
z_{12}z_{13}z_{14}z_{15}z_{23}z_{24}z_{26}z_{35}z_{36}
z_{45}z_{46}z_{56}
\end{gather} 
The corresponding graphs are, respectively,
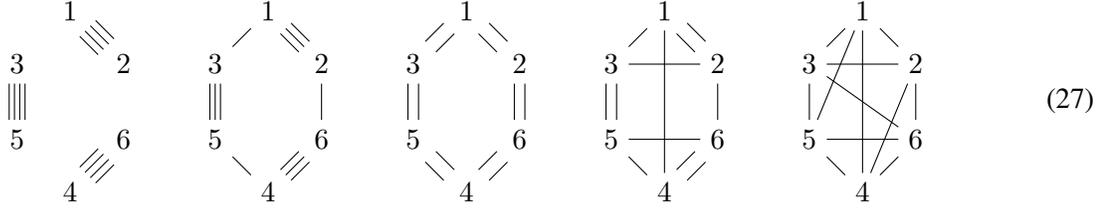
\begin{alignat}{5}
\begin{tikzpicture}[baseline={([yshift=-.5ex]current bounding box.center)}]
\node (v1) {$1$};
\node (v5) [below left of=v1] (v5) {$3$};
\node (v6) [below right of=v1] (v6) {$2$};
\node (v2) [below of=v5] (v2) {$5$};
\node (v3) [below of=v6] (v3) {$6$};
\node (v4) [below left of=v3] (v4) {$4$};
\draw[transform canvas={xshift=-3pt,yshift=-3pt}] (v1) -- (v6);
\draw[transform canvas={xshift=-1pt,yshift=-1pt}] (v1) -- (v6);
\draw[transform canvas={xshift=1pt,yshift=1pt}] (v1) -- (v6);
\draw[transform canvas={xshift=3pt,yshift=3pt}] (v1) -- (v6);
\draw[transform canvas={xshift=-3pt}] (v2) -- (v5);
\draw[transform canvas={xshift=-1pt}] (v2) -- (v5);
\draw[transform canvas={xshift=1pt}] (v2) -- (v5);
\draw[transform canvas={xshift=3pt}] (v2) -- (v5);
\draw[transform canvas={xshift=-3pt,yshift=3pt}] (v3) -- (v4);
\draw[transform canvas={xshift=-1pt,yshift=1pt}] (v3) -- (v4);
\draw[transform canvas={xshift=1pt,yshift=-1pt}] (v3) -- (v4);
\draw[transform canvas={xshift=3pt,yshift=-3pt}] (v3) -- (v4);
\end{tikzpicture}
\qquad
&
\begin{tikzpicture}[baseline={([yshift=-.5ex]current bounding box.center)}]
\node (v1) {$1$};
\node (v5) [below left of=v1] (v5) {$3$};
\node (v6) [below right of=v1] (v6) {$2$};
\node (v2) [below of=v5] (v2) {$5$};
\node (v3) [below of=v6] (v3) {$6$};
\node (v4) [below left of=v3] (v4) {$4$};
\draw (v1) -- (v6);
\draw[transform canvas={xshift=2pt,yshift=2pt}] (v1) -- (v6);
\draw[transform canvas={xshift=-2pt,yshift=-2pt}] (v1) -- (v6);
\draw (v3) -- (v4);
\draw[transform canvas={xshift=2pt,yshift=-2pt}] (v3) -- (v4);
\draw[transform canvas={xshift=-2pt,yshift=2pt}] (v3) -- (v4);
\draw (v2) -- (v5);
\draw[transform canvas={xshift=2pt}] (v2) -- (v5);
\draw[transform canvas={xshift=-2pt}] (v2) -- (v5);
\draw (v1)--(v5);
\draw (v2) -- (v4);
\draw (v3) -- (v6);
\end{tikzpicture}
\qquad
&
\begin{tikzpicture}[baseline={([yshift=-.5ex]current bounding box.center)}]
\node (v1) {$1$};
\node (v5) [below left of=v1] (v5) {$3$};
\node (v6) [below right of=v1] (v6) {$2$};
\node (v2) [below of=v5] (v2) {$5$};
\node (v3) [below of=v6] (v3) {$6$};
\node (v4) [below left of=v3] (v4) {$4$};
\draw[transform canvas={xshift=-2pt,yshift=2pt}] (v1) -- (v5);
\draw[transform canvas={xshift=2pt,yshift=-2pt}] (v1) -- (v5);
\draw[transform canvas={xshift=2pt,yshift=2pt}] (v1) -- (v6);
\draw[transform canvas={xshift=-2pt,yshift=-2pt}] (v1) -- (v6);
\draw[transform canvas={xshift=2pt,yshift=-2pt}] (v3) -- (v4);
\draw[transform canvas={xshift=-2pt,yshift=2pt}] (v3) -- (v4);
\draw[transform canvas={xshift=2pt}] (v3) -- (v6);
\draw[transform canvas={xshift=-2pt}] (v3) -- (v6);
\draw[transform canvas={xshift=2pt}] (v2) -- (v5);
\draw[transform canvas={xshift=-2pt}] (v2) -- (v5);
\draw[transform canvas={xshift=2pt,yshift=2pt}] (v2) -- (v4);
\draw[transform canvas={xshift=-2pt,yshift=-2pt}] (v2) -- (v4);
\end{tikzpicture}
\qquad
&
\begin{tikzpicture}[baseline={([yshift=-.5ex]current bounding box.center)}]
\node (v1) {$1$};
\node (v5) [below left of=v1] (v5) {$3$};
\node (v6) [below right of=v1] (v6) {$2$};
\node (v2) [below of=v5] (v2) {$5$};
\node (v3) [below of=v6] (v3) {$6$};
\node (v4) [below left of=v3] (v4) {$4$};
\draw[transform canvas={xshift=2pt,yshift=2pt}] (v1) -- (v6);
\draw[transform canvas={xshift=-2pt,yshift=-2pt}] (v1) -- (v6);
\draw[transform canvas={xshift=2pt,yshift=-2pt}] (v3) -- (v4);
\draw[transform canvas={xshift=-2pt,yshift=2pt}] (v3) -- (v4);
\draw[transform canvas={xshift=2pt}] (v2) -- (v5);
\draw[transform canvas={xshift=-2pt}] (v2) -- (v5);
\draw (v1) -- (v5);
\draw (v2) -- (v4);
\draw (v3) -- (v6);
\draw (v5) -- (v6);
\draw (v2) -- (v3);
\draw (v1) -- (v4);
\end{tikzpicture}
\qquad
&
\begin{tikzpicture}[baseline={([yshift=-.5ex]current bounding box.center)}]
\node (v1) {$1$};
\node (v5) [below left of=v1] (v5) {$3$};
\node (v6) [below right of=v1] (v6) {$2$};
\node (v2) [below of=v5] (v2) {$5$};
\node (v3) [below of=v6] (v3) {$6$};
\node (v4) [below left of=v3] (v4) {$4$};
\draw (v1) -- (v2);
\draw (v1) -- (v4);
\draw (v1) -- (v5);
\draw (v1) -- (v6);
\draw (v2) -- (v3);
\draw (v2) -- (v4);
\draw (v2) -- (v5);
\draw (v3) -- (v5);
\draw (v3) -- (v4);
\draw (v3) -- (v6);
\draw (v4) -- (v6);
\draw (v5) -- (v6);
\end{tikzpicture}
\end{alignat}
The symmetric polynomials can be written in terms of the elementary symmetric,
monomial symmetric, as well as the Schur polynomials, as before. We shall
abstain from writing all the
expressions since these are rather cumbersome. With $N=6$, $\delta=4$ and
$D=12$, it follows from \eq{gen:formula} that there can be at most
$18$ Schur polynomials. We can thus express
the five symmetric polynomials as linear combinations of these $18$ Schur
polynomials. The rank of this $5\times 18$ matrix turns out to be $2$. Using
the row reduction algorithm of \verb|Mathematica| we find three relations
among the five polynomials, 
\begin{gather}
S^{(6)}_1-4S^{(6)}_2=0,\\
S^{(6)}_4+2S^{(6)}_5=0,\\
S^{(6)}_2-S^{(6)}_3-S^{(6)}_4=0.
\end{gather} 
We can thus choose $S^{(6)}_3$ and $S^{(6)}_5$ as the independent
ones. We write the expressions for them below. 
\begin{equation}
\begin{split}
S^{(6)}_3&=\frac{1}{30}\left(
3 e_3^4-16 e_2 e_3^2 e_4+24 e_2^2 e_4^2-8 e_1 e_3 e_4^2+32
e_4^3-8 e_2^2 e_3 e_5+64 e_1 e_3^2 e_5-104 e_1 e_2 e_4 e_5\right.\\
&\left.\qquad-120 e_3 e_4 e_5+80 e_1^2 e_5^2+200 e_2 e_5^2+32 e_2^3 e_6
-120 e_1 e_2 e_3 e_6-24 e_3^2 e_6+200 e_1^2 e_4 e_6\right.\\
&\left.\qquad+304 e_2 e_4 e_6
-1960 e_1 e_5 e_6+5880 e_6^2\right)
\\
\ytableausetup{boxsize=3pt}
&=
\frac{1}{10}m_{\ydiagram{4,4,4}}
-\frac{2}{15}m_{\ydiagram{4,4,3,1}}
+\frac{1}{3}m_{\ydiagram{4,4,2,2}}
-\frac{2}{15}m_{\ydiagram{4,4,2,1,1}}
+\frac{4}{5}m_{\ydiagram{4,4,1,1,1,1}}
-\frac{2}{15}m_{\ydiagram{4,3,3,2}}\\
&\qquad+\frac{2}{3}m_{\ydiagram{4,3,3,1,1}}
-\frac{4}{15}m_{\ydiagram{4,3,2,2,1}}
-\frac{2}{5}m_{\ydiagram{4,3,2,1,1,1}}
+m_{\ydiagram{4,2,2,2,2}}  
+\frac{2}{5}m_{\ydiagram{4,2,2,2,1,1}}
+\frac{4}{5}m_{\ydiagram{3,3,3,3}} \\
&\qquad-\frac{2}{5}m_{\ydiagram{3,3,3,2,1}}
-\frac{32}{5}m_{\ydiagram{3,3,3,1,1,1}}
+\frac{2}{5}m_{\ydiagram{3,3,2,2,2}}
+\frac{56}{15}m_{\ydiagram{3,3,2,2,1,1}}
-\frac{44}{5}m_{\ydiagram{3,2,2,2,2,1}}
+66m_{\ydiagram{2,2,2,2,2,2}}
\\
&= \frac{1}{30}\big(
3 {s}_{(4,4,4)}-7 {s}_{(4,4,3,1)}+14 {s}_{(4,4,2,2)}-7
{s}_{(4,4,2,1,1)}+35 {s}_{(4,4,1,1,1,1)}-7 {s}_{(4,3,3,2)}\\
&\qquad+42{s}_{(4,3,3,1,1)}-35 {s}_{(4,3,2,2,1)}-56 {s}_{(4,3,2,1,1,1)}
+49{s}_{(4,2,2,2,2)}+98 {s}_{(4,2,2,2,1,1)}+35 {s}_{(3,3,3,3)}\\
&\qquad-56 {s}_{(3,3,3,2,1)}-154 {s}_{(3,3,3,1,1,1)}
+98 {s}_{(3,3,2,2,2)}+357{s}_{(3,3,2,2,1,1)}-1127 {s}_{(3,2,2,2,2,1)}\\
&\qquad+4508 {s}_{(2,2,2,2,2,2)}\big)
\end{split}
\end{equation} 
\begin{equation} 
\begin{split}
S^{(6)}_5&=\frac{1}{15}\left(
e_2^2 e_4^2-3 e_1 e_3 e_4^2+12 e_4^3-3 e_2^2 e_3 e_5+9 e_1 e_3^2 e_5+e_1 e_2
e_4 e_5-45 e_3 e_4 e_5-20 e_1^2 e_5^2+75 e_2 e_5^2 \right.\\ &\left.
\qquad+12 e_2^3 e_6-45 e_1 e_2 e_3 e_6+81 e_3^2 e_6+75 e_1^2 e_4 e_6
-126 e_2 e_4 e_6-135 e_1 e_5 e_6+405 e_6^2\right)\\
\ytableausetup{boxsize=3pt}
&=
\frac{1}{15}m_{\ydiagram{4,4,2,2}}
-\frac{1}{15}m_{\ydiagram{4,4,2,1,1}}
+\frac{2}{5}m_{\ydiagram{4,4,1,1,1,1}}
-\frac{1}{15}m_{\ydiagram{4,3,3,2}}
+\frac{1}{15}m_{\ydiagram{4,3,3,1,1}}
+\frac{1}{15}m_{\ydiagram{4,3,2,2,1}}
-\frac{1}{5}m_{\ydiagram{4,3,2,1,1,1}}
-\frac{2}{5}m_{\ydiagram{4,2,2,2,2}}  \\
&+\frac{1}{5}m_{\ydiagram{4,2,2,2,1,1}}
+\frac{2}{5}m_{\ydiagram{3,3,3,3}} 
-\frac{1}{5}m_{\ydiagram{3,3,3,2,1}}
+\frac{1}{5}m_{\ydiagram{3,3,2,2,2}}
+\frac{4}{15}m_{\ydiagram{3,3,2,2,1,1}}
-\frac{4}{5}m_{\ydiagram{3,2,2,2,2,1}}
+6m_{\ydiagram{2,2,2,2,2,2}}\\
&=\frac{1}{15}\big(
{s}_{(4,4,2,2)}-2 {s}_{(4,4,2,1,1)}+10 {s}_{(4,4,1,1,1,1)}
-2{s}_{(4,3,3,2)}+4 {s}_{(4,3,3,1,1)}+2 {s}_{(4,3,2,2,1)}\\
&\qquad-16{s}_{(4,3,2,1,1,1)}-14 {s}_{(4,2,2,2,2)}+28 {s}_{(4,2,2,2,1,1)}
+10{s}_{(3,3,3,3)}-16 {s}_{(3,3,3,2,1)}\\
&\qquad+28 {s}_{(3,3,3,1,1,1)}
+28{s}_{(3,3,2,2,2)}-6 {s}_{(3,3,2,2,1,1)}-70 {s}_{(3,2,2,2,2,1)}
+280{s}_{(2,2,2,2,2,2)}
\big)
\end{split}
\end{equation}
Here $S_3^{(6)}$ is a linear combination of all the 
$18$ Schur polynomials or $18$ monomial symmetric functions, 
but $S_5^{(6)}$ consists of only $16$ Schur polynomials or 
$15$ monomial symmetric functions.
Let us point out that the degree of the symmetric polynomials in terms of the
elementary symmetric polynomials is $\delta=4$, thus the
graphs are $4$-regular. 
\end{example}
\begin{example}
As the final example of the procedure let us work out the case of
$N=8$ particle. There are $11$ partitions, shown in Table~\ref{tab1}. 
We shall abstain from drawing the graphs in this case, which are all
$6$-regular and can be easily
drawn from the expressions of the graph monomials,
\begin{equation} 
\begin{split}
P_{{\mathcal G}_1} &= 
z_{12}^6z_{37}^6z_{46}^6z_{58}^6,\\
P_{{\mathcal G}_2} &= 
z_{12}^5z_{13}z_{28}z_{37}^5z_{46}^5z_{47}z_{56}z_{58}^5,\\
P_{{\mathcal G}_3} &= 
z_{12}^4z_{13}^2z_{28}^2z_{37}^4z_{46}^4z_{47}^2z_{56}^2z_{58}^4,\\
P_{{\mathcal G}_4} &= 
z_{12}^4z_{13}z_{14}z_{23}z_{28}z_{37}^4z_{46}^4z_{47}z_{56}z_{57}z_{58}^4z_{68},\\
P_{{\mathcal G}_5} &= 
z_{12}^3z_{13}^3z_{28}^3z_{37}^3z_{46}^3z_{47}^3z_{56}^3z_{58}^3,\\
P_{{\mathcal G}_6} &= 
z_{12}^3z_{13}^2z_{14}z_{23}z_{28}^2z_{37}^3z_{46}^3z_{47}^2z_{56}^2z_{57}z_{58}^3z_{68},\\
P_{{\mathcal G}_7} &= 
z_{12}^3z_{13}z_{14}z_{15}z_{23}z_{24}z_{28}z_{37}^3z_{38}z_{46}^3z_{47}
z_{56}z_{57}z_{58}^3z_{67}z_{68},\\
P_{{\mathcal G}_8} &= 
z_{12}^2z_{13}^2z_{14}^2z_{23}^2z_{28}^2z_{37}^2z_{46}^2z_{47}^2
z_{56}^2z_{57}^2z_{58}^2z_{68}^2,\\
P_{{\mathcal G}_9} &= 
z_{12}^2z_{13}^2z_{14}z_{15}z_{23}z_{24}z_{28}^2z_{37}^2z_{38}
z_{46}^2z_{47}^2z_{56}^2z_{57}z_{58}^2z_{67}z_{68},\\
P_{{\mathcal G}_{10}} &= 
z_{12}^2z_{13}z_{14}z_{15}z_{16}z_{23}z_{24}z_{25}z_{28}z_{34}z_{37}^2
z_{38}z_{46}^2z_{47}z_{56}z_{57}z_{58}^2z_{67}z_{68}z_{78},\\
P_{{\mathcal G}_{11}} &= 
z_{12}z_{13}z_{14}z_{15}z_{16}z_{17}z_{23}z_{24}z_{25}z_{26}z_{28}
z_{34}z_{35}z_{37}z_{38}z_{46}z_{47}z_{48}z_{56}z_{57}z_{58}z_{67}z_{68}z_{78}.
\end{split}
\end{equation} 
The corresponding eleven symmetric polynomials can be expressed, as before, in
terms of the elementary and monomial symmetric polynomials in turn. Then
these can be expressed in terms of Schur polynomials. Only $149$ among the 
$151$ Schur polynomials allowed by \eq{gen:formula}, with $N=6$, 
$\delta=6$ and $D=24$, occur in these expressions. 
The rank of the corresponding $11\times 149$ matrix turns out to be $4$.
Again, relations among the symmetric polynomials are obtained by row
reduction as
\begin{gather}
S^{(8)}_1 - 8S^{(8)}_2 = 0,\\ 
S^{(8)}_1-8S^{(8)}_5 -144 S^{(8)}_8+864 S^{(8)}_9 -612 S^{(8)}_{11}=0,\\
2S^{(8)}_3 -2S^{(8)}_5 -12 S^{(8)}_8 +40 S^{(8)}_9 -9S^{(8)}_{11}=0,\\ 
2S^{(8)}_6 + 4S^{(8)}_9 -3S^{(8)}_{11}=0,\\ 
2S^{(8)}_7 - 2S^{(8)}_9 +3S^{(8)}_{11}=0,\\
S^{(8)}_4 + 4S^{(8)}_7 =0,\\ 
2S^{(8)}_{10}- 3S^{(8)}_{11}=0. 
\end{gather}
We may take the independent ones as
$S^{(8)}_1$,
$S^{(8)}_5$,
$S^{(8)}_8$,
$S^{(8)}_{11}$.
\end{example}

{\it Conclusion:} We have described a method for construction of the ground state 
wave-function of a fractional quantum Hall system of $N$ interacting
electrons  under the influence of a magnetic flux $\Phi$ on the complex plane. 
The construction is very general. We assume that the
wave-function is antisymmetric as the electrons are shuffled and it furnishes 
a representation of the angular momentum algebra with total angular momentum
determined by the numbers of flux quanta and particles. 
Given only the number of electrons $N$ and the total integral flux 
$\Phi$, the procedure produces a set of 
linearly independent symmetric polynomials, satisfying these physical
assumptions. A particular linear combination
yields  the wave-function \eq{Psi:def} for a given interaction. 
These symmetric polynomials are, in turn, linear combinations of certain Schur 
polynomials and provide  the basis states in the occupation number 
representation, when multiplied by the Vandermonde polynomial.
By making the coordinates $z_i$ projective on a
sphere $\cp{1}$, the basis functions of the many-electron system on a
sphere can be constructed from the results described here.

{\it Acknowledgement:} SSM acknowledges support from SRIC, 
Indian Indian Institute of Technolgy, Kharagpur, under grant 
IIT/SRIC/PHY/EFH/2016-17/178.


\begin{thebibliography}{99}
\bibitem{GWW} M. Greiter, X.-G. Wen, and F. Wilczek, {\it Paired Hall state at half filling}, Phys. Rev. Lett. {\bf 66}, (1991) 3205.
\bibitem{RG} N. Read and D. Green, {\it Paired states of fermions in two dimensions with breaking of parity and time-reversal symmetries and the fractional quantum hall effect}, Phys. Rev. B {\bf 61}, (2000) 10267.
	
\bibitem{MR} G. Moore and N. Read, {\it Nonabelion in the fractional quantum Hall effect}, Nucl. Phys. B {\bf 360}, (1991) 362.
\bibitem{Jain} J. K. Jain, {\it Composite-Fermion approach for the fractional quantum hall effect}, Phys. Rev. lett. {\bf 63}, (1989) 199.
\bibitem{Wen} X. G. Wen, {\it Non-abelian statistics in the fractional quantum Hall states}, Phys. Rev. Lett. {\bf 66}, (1991) 802.
\bibitem{hald1} F. D. M. Haldane, \emph{Fractional Quantization of the Hall 
Effect:
A Hierarchy of Incompressible Quantum Fluid States}, Phys. Rev Lett. {\bf 51}
(1983) 605.
\bibitem{NW} C. Nayak and F. Wilczek, {\it $2n$-quasihole states realize $2^{n-1}$ dimensional spinor braiding statistics in paired quantum Hall states}, Nucl. Phys. B {\bf 479}, (1996) 529.
\bibitem{RR} N. Read and R. Rezayi, {\it Beyond paired quantum hall states: Parefermions and incompressible states in the first excited Landau level}, Phys. Rev. B {\bf 59}, (1999) 8084.
\bibitem{hald2} B. Andrei Bernevig and F. D. M. Haldane, 
\emph{Model Fractional Quantum Hall States and Jack Polynomials},
Phys. Rev. Lett. {\bf 100} (2008) 246802.
\bibitem{Rezayi}V. W. Scarola, J. K. Jain, E. H. Rezayi, {\it Pairing-induced even denominator fractional quantum hall effect in the lowest landau level}, Phys. Rev. Lett. {\bf 88}, (2002) 216804.
\bibitem{Macdonald} I. G. Macdonald, {\it Symmetric Functions and Hall
	Polynomials,\/}
Oxford University Press, Oxford, 1979.

\bibitem{GS} G.~Sabidussi, \emph{Binary invariants and orientations of
graphs}, Discrete Mathematics {\bf 101} (1992) 251.

\bibitem{vakil} B.~Howard, J.~Millson, A.~Snowden, R.~Vakil,
\emph{The equations for the moduli space of $n$ points on the line}
Duke Math. J.\ {\bf 146}(2009)175.

\bibitem{Li}{C.~L.~S.~Batista and D.~Li}, \emph{Analytic calculations of trial
wave functions of the fractional quantum Hall effect on the sphere}, Phys.
Rev. {\bf B55}(1997)1582.

\bibitem{M2}
D.~R.~Grayson and M.~E.~Stillman,
\verb|Macaulay2|, \emph{a software system for research in algebraic
geometry}, {Available at {http://www.math.uiuc.edu/Macaulay2/} }
\bibitem{stone}
M.~Stone, \emph{Schur functions, chiral bosons, and the quantum-Hall-effect edge
states}, Phys. Rev. {\bf B42} (1990) 8399.

\end{thebibliography}
\end{document}